\documentclass{article}

\newcommand{\Di}{\displaystyle}

\begin{document}

\title{A new strategy to find bound states in anharmonic oscillators}

\author{Francisco J. G\'omez and Javier Sesma\\  \   \\
Departamento de F\'{\i}sica Te\'{o}rica, \\ Facultad de Ciencias,
\\ 50009 Zaragoza, Spain. \\  \ }

\maketitle

\begin{abstract}
A very simple procedure to calculate eigenenergies of quantum
anharmonic oscillators is presented. The method, exact but for
numerical computations, consists merely in requiring the vanishing
of the Wronskian of two solutions which are regular respectively
at the origin and at infinity. The first one can be represented by
its series expansion; for the second one, an asymptotic expansion
is available. The procedure is illustrated by application to
quartic and sextic oscillators.
\end{abstract}

Anharmonic oscillators have been frequently used to modelize
different problems in all branches of Physics. In fact, anharmonic
terms are unavoidable to represent a realistic potential in the
vicinity of a local minimum. Besides this, since the seminal paper
by Bender and Wu on the quartic oscillator \cite{bend}, plenty of
mathematical questions having anharmonic oscillators as their
subject have risen. Consequently, a great diversity of
(perturbative, iterative, variational, etc.) methods have been
proposed to find the eigenvalues of the Hamiltonian. Here we
consider an isotropic D-dimensional anharmonic oscillator and
present a very simple method that provides an exact quantization
condition: the eigenvalues turn out to be the zeros of an easily
obtainable function of the energy. Of course, the determination of
those zeros cannot be done without having recourse to numerical
computation, but the accuracy of the results can be controlled.

Let us consider the Schr\"odinger equation (in units
$\hbar^2/2m=1$)
\begin{equation}
-\,u^{\prime\prime}(r) + V(r)\,u(r) = E\,u(r),  \label{uno}
\end{equation}
with an effective (including centrifugal terms) anharmonic
potential,
\begin{equation}
V(r)= \sum_{j=-2}^{2N}A_j\,r^j, \qquad N\geq 2, \qquad  A_{2N}>0,
\label{dos}
\end{equation}
all parameters being assumed real. Such a potential may correspond
either to a one-dimensional oscillator, with $-\infty < r <
+\infty$ and $\int_{-\infty}^{+\infty}|u|^2dr=1$, or to an
isotropic $D$-dimensional ($D>1$) oscillator, with $0\leq r <
+\infty$ and $\int_0^{+\infty}|u|^2dr=1$. In the second case, the
reduced radial wave function must satisfy the condition $u(0)=0$.
Two independent power series solutions of Eq. (\ref{uno}), of the
form
\begin{equation}
u(r)=\sum_{n=0}^\infty a_n\,r^{n+\nu},\qquad a_0\neq 0,
\label{tres}
\end{equation}
with
\begin{equation}
\nu = \frac{1}{2}\left(1\pm\sqrt{1+4A_{-2}}\right),  \label{ocho}
\end{equation}
can be immediately obtained by direct substitution. But it may
occur, depending on the values of $D$ and $A_{-2}$, that only one
of such mathematical solutions, $u_{\scriptstyle {\rm {reg}}}(r)$,
satisfies the requirement of regularity at $r=0$. The
normalizability of the wavefunction imposes a regularity at
infinity that takes place only if the energy $E$ adopts one of the
eigenvalues $E_k$, $k=0, 1, 2, \ldots$. Asymptotic expansions of
the wavefunction for large $r$, like
\begin{equation}
u(r)\sim\exp\left[\sum_{p=1}^{N+1}\frac{\alpha_p}{p}\,r^p\right]
r^{\mu}\sum_{m=0}^\infty h_m\,r^{-m}, \qquad h_0\neq 0,
\label{cuatro}
\end{equation}
can also be found by substitution. In fact, two independent
solutions, $u^{(1)}(r)$ and $u^{(2)}(r)$, exist having asymptotic
expansions of the form (\ref{cuatro}). In what follows, we will
assume that $u^{(1)}(r)$ represents the solution well behaved for
$r\to\infty$ and $u^{(2)}(r)$ a divergent one. Obviously,
$u_{\scriptstyle {\rm {reg}}}$ can be written as a linear
combination of $u^{(1)}$ and $u^{(2)}$ with coefficients $T^{(1)}$
and $T^{(2)}$, usually called {\em connection factors}, which
depend on the parameters $A_j$ of the potential and on the energy
$E$. For a given set of potential parameters, the eigenvalues
$E_k$ turn out to be the zeros of $T^{(2)}$ as a function of $E$.

Explicit expressions of $T^{(2)}$ in terms of $E$ can be obtained
only in some particular cases. Nevertheless, $T^{(2)}$ can be
written as the quotient of two (independent of $r$) Wronskians,
\begin{equation}
T^{(2)}=\mathcal{W}[u_{\scriptstyle {\rm
{reg}}},u^{(1)}]/\mathcal{W}[u^{(2)},u^{(1)}].  \label{cinco}
\end{equation}
As far as we are not interested in the precise value of $T^{(2)}$
but in their zeros as a function of $E$, the determination of the
eigenvalues reduces to find the values of $E$ for which the
Wronskian
\begin{equation}
\mathcal{W}[u_{\scriptstyle {\rm {reg}}},u^{(1)}]=0.  \label{seis}
\end{equation}
As it can be seen, this quantization condition is most simple. Its
implementation, however, requires certain tricks, because of the
formal nature of the expansion (\ref{cuatro}). We are going to
illustrate the procedure in two particular cases that have
received considerable attention: the quartic anharmonic oscillator
and the sextic one.

The quartic oscillator, corresponding to take
\begin{equation}
V(r)=A_4\,r^4+A_2\,r^2+ A_{-2}\,r^{-2}, \qquad A_4>0,
\label{siete}
\end{equation}
in (\ref{dos}), is the best known anharmonic oscillator. Its
one-dimensional version has served as test bed for different
approximate methods of solution of the Schr\"odinger equation. For
the asymptotic expansions of the form (\ref{cuatro}) we have the
two sets of exponents
\begin{equation}
\begin{array}{llll}
\alpha_3^{(1)}=-\sqrt{A_4}, \quad & \alpha_2^{(1)}=0, \quad &
\alpha_1^{(1)}=-A_2/2\sqrt{A_4}, \quad & \mu^{(1)} =-1, \\
\alpha_3^{(2)}=\sqrt{A_4}, \quad & \alpha_2^{(2)}=0, \quad &
\alpha_1^{(2)}=A_2/2\sqrt{A_4}, \quad & \mu^{(2)} =-1,
\end{array}  \label{diez}
\end{equation}
and the recurrence
\begin{equation}
2\alpha_3mh_m=(E+\alpha_1^2)h_{m-1}-2\alpha_1(m-1)h_{m-2}+
((m-1)(m-2)-A_{-2})h_{m-3}.  \label{duno}
\end{equation}
The power series for $u_{\scriptstyle {\rm {reg}}}$ and the
asymptotic expansion for $u^{(1)}$ would allow one to obtain a
formal expression for the Wronskian in (\ref{seis}). For technical
reasons, we prefer introducing two auxiliary functions
\begin{equation}
v_{\scriptstyle {\rm
{reg}}}(r)=\exp(-\alpha_3^{(1)}r^3/3)\,u_{\scriptstyle {\rm
{reg}}}(r), \qquad
v^{(1)}(r)=\exp(-\alpha_3^{(1)}r^3/3)\,u^{(1)}(r), \label{ddos}
\end{equation}
whose Wronskian turns out to be
\begin{equation}
\mathcal{W}[v_{\scriptstyle {\rm
{reg}}},v^{(1)}]=\exp(-2\alpha_3^{(1)}r^3/3)\,
\mathcal{W}[u_{\scriptstyle {\rm {reg}}},u^{(1)}].  \label{dtres}
\end{equation}
By substituting $v^{(1)}$ by its asymptotic expansion and denoting
\begin{equation}
w(r)=\exp(\alpha_1^{(1)}r)\, v_{\scriptstyle {\rm {reg}}}(r),
\label{dcuatro}
\end{equation}
we obtain for the left hand side of (\ref{dtres})
\begin{eqnarray}
\lefteqn{\hspace{-1cm} \mathcal{W}[v_{\scriptstyle {\rm
{reg}}},v^{(1)}]\sim \left(
2\alpha_1^{(1)}w(r)-w^{\prime}(r)\right) \left(
\sum_{m=0}^{\infty}h_m^{(1)}r^{-m-1}\right)} \nonumber \\ & &
\hspace{4cm}
+\,w(r)\left(\sum_{m=0}^{\infty}h_m^{(1)}r^{-m-1}\right)^{\prime}.
\label{dcinco}
\end{eqnarray}
The power series expansion
\begin{equation}
w(r)=\sum_{n=0}^\infty b_n\,r^{n+\nu}, \qquad b_0\neq 0,
\label{dseis}
\end{equation}
can be easily obtained as solution of the differential equation
\begin{equation}
w^{\prime\prime}+2(\alpha_3^{(1)}r^2-\alpha_1^{(1)})w^{\prime} +
(-2A_2r^2+2\alpha_3^{(1)}r+E+(\alpha_1^{(1)})^2-A_{-2}r^{-2})w=0.
\label{dsiete}
\end{equation}
The recurrence obeyed by the coefficients is
\begin{eqnarray}
\lefteqn{ \hspace{-1cm}n(n-1+2\nu )b_n=2\alpha_1^{(1)}(n-1+\nu
)b_{n-1} -(E+(\alpha_1^{(1)})^2)b_{n-2}} \nonumber  \\ & &
\hspace{3cm} -\,2\alpha_3^{(1)}(n-2+\nu )b_{n-3} +2A_2b_{n-4}.
\label{docho}
\end{eqnarray}
By choosing, for instance, the starting values
\begin{equation}
h_0^{(1)}=1 \quad {\rm {and}} \quad b_0=1,  \label{dnueve}
\end{equation}
it is immediate to obtain, by means of the recurrences
(\ref{duno}) and (\ref{docho}), an explicit form for the
asymptotic expansion (\ref{dcinco}), that, with an obvious
notation, would read
\begin{equation}
\mathcal{W}[v_{\scriptstyle {\rm {reg}}},v^{(1)}]\sim
\sum_{k=-\infty}^{\infty}\gamma_k\,r^{k-1+\nu}.  \label{veinte}
\end{equation}
According to (\ref{dtres}), the last expression must be a formal
expansion of
\begin{equation}
\exp(-2\alpha_3^{(1)}r^3/3) \, \mathcal{W}[u_{\scriptstyle {\rm
{reg}}},u^{(1)}],  \label{vuno}
\end{equation}
the Wronskian of $u_{\scriptstyle {\rm {reg}}}$ and $u^{(1)}$
being independent of $r$. The crucial point of our procedure is to
recall the Heaviside's exponential series
\begin{equation}
\exp(t)\sim\sum_{n=-\infty}^{\infty}\frac{t^{n+\delta}}{\Gamma(n+1+\delta)},
\label{vdos}
\end{equation}
introduced by Heaviside in the second volume of his {\em
Electromagnetic theory} (London, 1899), probed by Barnes
\cite{barn} to be an asymptotic expansion for arbitrary $\delta$
and $|\arg (t)|<\pi$, and used by Naundorf \cite{naun} in his
treatment of the connection problem, which has provided the
inspiration for our procedure. An asymptotic expansion for the
expression (\ref{vuno}) can be immediately written. It becomes
\begin{equation}
\exp(-2\alpha_3^{(1)}r^3/3) \,\mathcal{W}[u_{\scriptstyle {\rm
{reg}}},u^{(1)}]\sim \beta_1\mathcal{E}_1
+\beta_2\mathcal{E}_2+\beta_3\mathcal{E}_3, \label{vtres}
\end{equation}
where
\begin{eqnarray}
\mathcal{E}_1&\sim&\sum_{n=-\infty}^{\infty}
\frac{((-2/3)\alpha_3^{(1)}r^3)^{n+\nu/3}}{\Gamma(n+1+\nu/3)},  \\
\mathcal{E}_2&\sim&\sum_{n=-\infty}^{\infty}
\frac{((-2/3)\alpha_3^{(1)}r^3)^{n+(\nu+1)/3}}{\Gamma(n+1+(\nu+1)/3)},
\\  \mathcal{E}_3&\sim&\sum_{n=-\infty}^{\infty}
\frac{((-2/3)\alpha_3^{(1)}r^3)^{n+(\nu+2)/3}}{\Gamma(n+1+(\nu+2)/3)},
\end{eqnarray}
and the constants $\beta_1$, $\beta_2$ and $\beta_3$ are
arbitrary, with the only restriction
\begin{equation}
\mathcal{W}[u_{\scriptstyle {\rm
{reg}}},u^{(1)}]=\beta_1+\beta_2+\beta_3. \label{vsiete}
\end{equation}
Identification of the right hand sides of (\ref{veinte}) and
(\ref{vtres}) allows one to obtain the values of $\beta_1$,
$\beta_2$ and $\beta_3$, that, substituted in (\ref{vsiete}), give
\begin{eqnarray}
\lefteqn{\hspace{-0.5 cm}\mathcal{W}[u_{\scriptstyle {\rm
{reg}}},u^{(1)}]=\frac{1}{((-2/3)\alpha_3^{(1)})^{n+\nu/3}}
\,\Bigg( \Gamma(n+1+\nu/3)\,\gamma_{3n+1}  \qquad\qquad} \nonumber
\\ & & \hspace{0.5 cm}+ \, \frac{\Gamma(n+1+(\nu+1)/3)}
{((-2/3)\alpha_3^{(1)})^{1/3}}\,\gamma_{3n+2}+\frac{\Gamma(n+1+(\nu+2)/3)}
{((-2/3)\alpha_3^{(1)})^{2/3}}\,\gamma_{3n+3}\Bigg), \label{vocho}
\end{eqnarray}
where the index $n$ can be chosen at will. For positive values of
the subindex $k$, the coefficients of the expansion (\ref{veinte})
are given by
\begin{equation}
\gamma_k=\sum_{m=0}^{\infty}h_m^{(1)}\left(
2\alpha_1^{(1)}\,b_{k+m} -(2m+k+2+\nu)\,b_{k+m+1}\right).
\label{vnueve}
\end{equation}

In order to compare with reliable published results, we have
applied our procedure to the one-dimensional potential
\begin{equation}
V(r)=r^4+A_2r^2,  \label{tcero}
\end{equation}
discused by Balsa {\em et al.} \cite{bals}. Their results for
$A_2<0$ are quoted in all posterior papers dealing with different
approximate methods to solve the double-well anharmonic
oscillator. Table 1 shows the four lowest eigenenergies, for
several values of $A_2$, obtained by requiring the cancellation of
the right hand side of (\ref{vocho}). In the computation we have
used a FORTRAN program with double precision. Our results are in
perfect coincidence with those for $E_0$ and $E_1$ given in Ref.
\cite{bals}.

\begin{table}
\begin{tabular}{rrrrr}
\hline $A_2$ & $E_0$\hspace{0.8 cm} & $E_1$\hspace{0.8 cm}  &
$E_2$\hspace{0.8 cm} & $E_3$\hspace{0.8 cm} \\
\hline 0 & \qquad 1.06036209 & \quad 3.79967303 &
\quad 7.45569794 & \quad 11.64474551 \\
--\,1 & \qquad 0.65765301 & \quad 2.83453620 & \quad 6.16390126 &
\quad 10.03864612 \\
--\,2 & \qquad 0.13778585 & \quad 1.71302790 & \quad 4.78242971 &
\quad 8.33286819 \\
--\,3 & \qquad --\,0.59349330 & \quad 0.37766207 & \quad
3.34533567 & \quad 6.52498666 \\
--\,4 & \qquad --\,1.71035045 & \quad --\,1.24792249 & \quad
1.94143719 & \quad 4.61294345 \\
--\,5 & \qquad --\,3.41014276 & \quad --\,3.25067536 & \quad
0.63891956 & \quad 2.58121627 \\
--\,6 & \qquad --\,5.74819052 & \quad --\,5.70679252 & \quad
--\,0.72394168 & \quad 0.37528499 \\
--\,7 & \qquad --\,8.67110521 & \quad --\,8.66245222 & \quad
--\,2.54370521 & \quad --\,2.11199938 \\
--\,8 & \qquad --\,12.13633072 & \quad --\,12.13481435 & \quad
--\,5.12655020 & \quad --\,5.01091331 \\
--\,9 & \qquad --\,16.12618646 & \quad --\,16.12595855 & \quad
--\,8.44212291 & \quad --\,8.41871412 \\
--\,10 & \qquad --\,20.63357670 & \quad --\,20.63354688 & \quad
--\,12.37954379 & \quad --\,12.37567372 \\ \hline
\end{tabular}
\caption{The four lowest eigenenergies of the double-well
potential (\ref{tcero}), for several values of $A_2$. The energies
$E_0$ and $E_2$ correspond to even states ($\nu=0$ in
(\ref{ocho})); $E_1$ and $E_3$ to odd ones ($\nu=1$).}
\end{table}

The second example of the application of our strategy refers to
the sextic anharmonic potential
\begin{equation}
V(r)=A_6\,r^6+A_4\,r^4+A_2\,r^2+A_{-2}\,r^{-2}, \qquad A_6>0,
\label{treinta}
\end{equation}
which, besides having served as a model for a great variety of
systems, like interacting anyons, molecules of ammonia, and
hydrogen-bonded solids, is a well known example of quasi-exactly
solvable (QES) potential for certain values of the parameters. For
 the asymptotic expansions
(\ref{cuatro}), the non-vanishing exponents are now
\begin{equation}
\begin{array}{lll}
\alpha_4^{(1)}=-\sqrt{A_6}, \quad & \alpha_2^{(1)}=\frac{\Di
-A_4}{\Di 2\sqrt{A_6}}, \quad  & \mu^{(1)} = -\frac{\Di 3}{\Di 2}-
\frac{\Di 4A_6A_2-A_4^2}{\Di 8A_6\sqrt{A_6}}, \\
\alpha_4^{(2)}=\sqrt{A_6}, \quad & \alpha_2^{(2)}=\frac{\Di
A_4}{\Di 2\sqrt{A_6}}, \quad & \mu^{(2)} =-\frac{\Di 3}{\Di
2}+\frac{\Di 4A_6A_2-A_4^2}{\Di 8A_6\sqrt{A_6}},
\end{array}  \label{tdos}
\end{equation}
and the recurrence
\begin{eqnarray}
\lefteqn{\hspace{-1cm}2\alpha_4mh_m=(E+\alpha_2(-2m+5+2\mu))h_{m-2}}
\nonumber
\\ & & \hspace{3cm} + \, ((m-4-\mu)(m-3-\mu)-A_{-2})h_{m-4}.  \label{ttres}
\end{eqnarray}
As auxiliary functions we choose now
\begin{equation}
v_{\scriptstyle {\rm
{reg}}}(r)=\exp(-\alpha_4^{(1)}r^4/4)\,u_{\scriptstyle {\rm
{reg}}}(r), \qquad
v^{(1)}(r)=\exp(-\alpha_4^{(1)}r^4/4)\,u^{(1)}(r), \label{tcuatro}
\end{equation}
whose  Wronskian verifies
\begin{equation}
\mathcal{W}[v_{\scriptstyle {\rm
{reg}}},v^{(1)}]=\exp(-\alpha_4^{(1)}r^4/2)\,
\mathcal{W}[u_{\scriptstyle {\rm {reg}}},u^{(1)}].  \label{tcinco}
\end{equation}
By substituting $v^{(1)}$ by its asymptotic expansion and denoting
\begin{equation}
w(r)=\exp(\alpha_2^{(1)}r^2/2)\, v_{\scriptstyle {\rm {reg}}}(r),
\label{tseis}
\end{equation}
we obtain for the left hand side of (\ref{tcinco})
\begin{eqnarray}
\lefteqn{\hspace{-1cm}\mathcal{W}[v_{\scriptstyle {\rm
{reg}}},v^{(1)}]\sim \left(
2\alpha_2^{(1)}rw(r)-w^{\prime}(r)\right)
\left(\sum_{m=0}^{\infty}h_m^{(1)}r^{-m+\mu^{(1)}}\right)}
\nonumber \\ & & \hspace{4.5 cm} + \, w(r)\left(
\sum_{m=0}^{\infty}h_m^{(1)}r^{-m+\mu^{(1)}}\right)^{\prime}.
\label{tsiete}
\end{eqnarray}
Since $w(r)$ obeys the differential equation
\begin{eqnarray}
\lefteqn{\hspace{-0.5cm} w^{\prime\prime}+
2(\alpha_4^{(1)}r^3-\alpha_2^{(1)}r)w^{\prime}} \nonumber  \\  & &
\hspace{-0.5 cm} +\, (-2A_4r^4+(3\alpha_4^{(1)}+(\alpha_2^{(1)})^2
-A_2)r^2 +E-\alpha_2^{(1)}-A_{-2}r^{-2})w=0, \label{tocho}
\end{eqnarray}
the coefficients of its series expansion
\begin{equation}
w(r)=\sum_{n=0}^\infty b_n\,r^{n+\nu}, \qquad b_0\neq 0,
\label{tnueve}
\end{equation}
can be easily obtained by means of the recurrence
\begin{eqnarray}
\lefteqn{\hspace{-1cm}n(n-1+2\nu )b_n=
(-E+\alpha_2^{(1)}(2n-3+2\nu))b_{n-2}} \nonumber \\ & &
\hspace{1cm}+ \, (A_2-(\alpha_2^{(1)})^2
-\alpha_4^{(1)}(2n-5+2\nu))b_{n-4} +\, 2A_4b_{n-6}.
\label{cuarenta}
\end{eqnarray}
The expansion (\ref{tsiete}) adopts the form
\begin{equation}
\mathcal{W}[v_{\scriptstyle {\rm {reg}}},v^{(1)}]\sim
\sum_{k=-\infty}^{\infty}\gamma_{2k}\,r^{2k+1+\nu+\mu^{(1)}},
\label{cuno}
\end{equation}
where we have denoted
\begin{equation}
\gamma_{2k}=\sum_{m=0}^{\infty}h_{2m}^{(1)}\left(
2\alpha_2^{(1)}\,b_{2k+2m}
-(2k+4m+2+\nu-\mu^{(1)})\,b_{2k+2m+2}\right). \label{cdos}
\end{equation}
The right hand side of (\ref{tcinco}) can now be written as
\begin{equation}
\exp(-\alpha_4^{(1)}r^4/2) \,\mathcal{W}[u_{\scriptstyle {\rm
{reg}}},u^{(1)}]\sim \beta_1\mathcal{E}_1 +\beta_2\mathcal{E}_2,
\label{ctres}
\end{equation}
with
\begin{eqnarray}
\mathcal{E}_1&\sim&\sum_{n=-\infty}^{\infty}
\frac{((-1/2)\alpha_4^{(1)}r^4)^{n+(1+\nu+\mu^{(1)})/4}}
{\Gamma(n+1+(1+\nu+\mu^{(1)})/4)},  \\
\mathcal{E}_2&\sim&\sum_{n=-\infty}^{\infty}
\frac{((-1/2)\alpha_4^{(1)}r^4)^{n+(3+\nu+\mu^{(1)})/4}}
{\Gamma(n+1+(3+\nu+\mu^{(1)})/4)},
\end{eqnarray}
and the constants $\beta_1$ and $\beta_2$ being such that
\begin{equation}
\mathcal{W}[u_{\scriptstyle {\rm {reg}}},u^{(1)}]=\beta_1+\beta_2.
\label{cseis}
\end{equation}
Comparison of the right hand sides of (\ref{cuno}) and
(\ref{ctres}) gives the values of $\beta_1$ and $\beta_2$ and,
finally,
\begin{eqnarray}
\lefteqn{ \hspace{-0.5cm}\mathcal{W}[u_{\scriptstyle {\rm
{reg}}},u^{(1)}]=\frac{1}{(-\alpha_4^{(1)}/2)^{n+(1+\nu+\mu^{(1)}))/4}}
\,\Bigg( \Gamma(n+1+(1+\nu+\mu^{(1)})/4)\,\gamma_{4n}
\qquad\qquad} \nonumber
\\ & & \hspace{4cm}\ +\frac{\Gamma(n+1+(3+\nu+\mu^{(1)})/4)}
{(-\alpha_4^{(1)}/2)^{1/2}}\,\gamma_{4n+2}\Bigg), \label{csiete}
\end{eqnarray}
where the index $n$ is any integer.

We have looked for the values of $E$ which make the right hand
side of (\ref{csiete}) to become zero in the case of a potential
of the form (\ref{treinta}), namely
\begin{equation}
V(r)=r^6-(4s+4J-2)\,r^2+\frac{1}{4}(4s-1)(4s-3)\,r^{-2}.
\label{cocho}
\end{equation}
Such a potential was first discussed by Turbiner \cite{turb} and
has been also considered by Bender and Dunne \cite{dunn} and by
Finkel {\em et al.} \cite{fink} to illustrate the existence of a
family of orthogonal polynomials associated to the quasi-exact
solvability. Whenever the parameter $J$ takes positive integer
values, that potential becomes QES for any value of $s$. In fact,
$J$ is the number of eigenvalues and eigenfunctions that can be
obtained exactly. In Table 2 we report the results of our
procedure by calculating the lowest eigenenergies for a particular
value of $s$ and different values of $J$. For $J=1, 2, 3, 4$, the
$J$ lowest values of the energy coincide exactly with the zeros of
the corresponding critical polynomials of Bender and Dunne. In
these cases, the coefficients $h_m$ vanish for $m$ larger than a
certain $M$, depending on $J$, and, consequently, the expansion
(\ref{cuatro}) becomes exact and gives the quasi-exact
eigenfunction.

\begin{table}
\begin{tabular}{crrrr}
\hline $J$ & $E_0$\hspace{0.8 cm} & $E_1$\hspace{0.8 cm}  &
$E_2$\hspace{0.8 cm} & $E_3$\hspace{0.8 cm} \\
\hline $-\sqrt{3}/4$ & \qquad 5.75218468 & \quad 17.19822587 &
\quad 32.16502803 & \quad 49.95708442 \\
0 & \qquad 4.22831744 & \quad 15.21728994 & \quad 29.75575600 &
\quad 47.18056681 \\
0.5 & \qquad 2.25118051 & \quad 12.87099068 & \quad 26.92815514 &
\quad 43.93525080 \\
1 & \qquad 0.00000000 & \quad 10.46738163 & \quad 24.05793091 &
\quad 40.65160026 \\
1.5 & \qquad --\,2.56314639 & \quad 8.00423193 & \quad 21.15280932
& \quad 37.33449367 \\
2 & \qquad --\,5.46410162 & \quad 5.46410162 & \quad 18.22120991 &
\quad 33.98950899 \\
2.5 & \qquad --\,8.71290067 & \quad 2.81198082 & \quad 15.27080313
& \quad 30.62297181 \\
3 & \qquad --\,12.30551201 & \quad 0.00000000 & \quad 12.30551201
& \quad 27.24194193 \\
3.5 & \qquad --\,16.22899813 & \quad --\,3.02264936 & \quad
9.32087564 & \quad 23.85400371 \\
4 & \qquad --\,20.46665929 & \quad --\,6.29920011 & \quad
6.29920011 & \quad 20.46665929 \\
\hline
\end{tabular}
\caption{The four lowest eigenenergies of the potential
(\ref{cocho}) with $s=(2+\sqrt{3})/4$ and different values of $J$.
Positive integer values of $J$ give QES potentials. The value
$J=-\sqrt{3}/4$ corresponds to an effective potential
$r^6+(1/2)r^{-2}$.}
\end{table}

The above described method of solving the Schr\"odinger equation
with anharmonic potentials is exact, in principle, but its
usefulness lies on two assumptions: the convergence of series like
those in (\ref{vnueve}) and (\ref{cdos}) and the numerical
stability of recurrences like (\ref{duno}), (\ref{docho}),
(\ref{ttres}) and (\ref{cuarenta}), in the examples considered.
Investigations tending to justify both assumptions are in
progress.

It is a great pleasure to dedicate this contribution to Prof.
Alberto Galindo on occasion of his seventieth birthday. Financial
support from Comisi\'{o}n Inter\-mi\-nis\-te\-rial de Ciencia y
Tecnolog\'{\i}a is acknowledged.

\end{document}